\documentclass[preprint,12pt, authoryear]{elsarticle}
\usepackage[top=1in, bottom=1in, left=1in, right=1in]{geometry}
\usepackage{bm}
\usepackage{url}
\usepackage{float}
\usepackage{color}
\usepackage{wrapfig}
\usepackage[ruled, vlined]{algorithm2e}
\usepackage{longtable,booktabs, makecell}
\usepackage{amsfonts,mathrsfs,amsthm,amssymb,amsmath}
\usepackage{multirow}
\usepackage{tabularx}
\usepackage{adjustbox}
\usepackage{newtxtext,newtxmath}
\usepackage{algorithm2e}
\usepackage{setspace}
\usepackage{graphicx}
\usepackage[colorlinks=true, linkcolor=blue, citecolor=red, urlcolor=magenta]{hyperref}
\newtheorem{prop}{\sc Proposition}[section]
\newtheorem{coro}{\sc Corollary}[section]

\newtheorem{theorem}{\sc Theorem}[section]

\journal{Applied Mathematical Modelling}

\begin{document}

\begin{frontmatter}

\title{\textbf{Adaptive weighted approach for high-dimensional statistical learning and inference}}

\author{Jun Lu$^1$, Xiaoyu Ma$^1$, Mengyao Li$^2$ and Chenping Hou$^1$\corref{mycorrespondingauthor}}
\address{$^1$National University of Defense Technology, Changsha, Hunan, China\\$^2$Xi'an Jiaotong University, Xi'an, Shanxi, China}

\cortext[mycorrespondingauthor]{The Corresponding author(hcpnudt@hotmail.com). The first two authors contributed equally.}

\begin{abstract}
We propose a new weighted average estimator for the high dimensional parameters under the distributed learning system, in which the weight assigned to each coordinate is precisely proportional to the inverse of the variance of the local estimates for that coordinate. 
This strategy empowers the new estimator to achieve a minimal mean squared error, comparable to the current state-of-the-art one-shot distributed learning methods.
While at the same time, the new weighting approach maintains remarkably low communication costs, as each agent is required to transmit only two vectors to the central server. 
As a result, the newly proposed method achieves optimal statistical efficiency while significantly reducing communication overhead.
We further demonstrate the effectiveness of the new estimator by investigating the error bound and the asymptotic properties of the estimation, as well as the numerical performance on some simulated examples and a real data analysis.
\end{abstract}


\begin{keyword}
	distributed learning \sep high-dimensional parameter \sep statistical efficient \sep low communication cost
\end{keyword}
\end{frontmatter}


\section{Introduction}

The distributed learning (DL) has attracted much attention in diverse fields due to its high efficiency in dealing with distributed. 
In brief, DL adopts a master-and-agents distributed architecture, in which the agents distributed over the network are like ``intelligent workers''  having the ability to execute some calculations or perform some tasks, and the master is like an ``leader'' located in the central position in the network and can communicate with the agents \cite{mo2024distributed}. 
DL usually takes the following workflow: the agents first learn some ``local'' results in a parallel way, then the master receives the local results and ensembles a global one.
DL is also a powerful strategy to deal with the large-scale dataset, particularly for those whose size exceed the memory capacity of personal computers. By splitting the massive dataset into many smaller pieces, even a standard computer can easily handle this type of dataset.
%

There is a large body of literature on DL, which can generally be categorized into two approaches: the one-shot learning (OSL) and the multiple-shot learning (MSL). 
In the case of OSL, each agent is required to communicate with the central server only once to transmit the local results. This approach is particular well-suited for the scenarios where the communication costs are high. 
Notable contributions on OSL for high-dimensional parameter include the works by \cite{lee2017communication}, \cite{tang2020distributed} and \cite{lv2022debiased}, which focus on the statistical models such as generalized linear models and partially linear models with sparse parameter.
\cite{liu2024statistical} studied the statistical optimality of divide and conquer kernel-based functional linear regression.
\cite{fan2024distributed} discussed the distributed quantile regression method for longitudinal big data. 
All the aforementioned studies employed the simple average strategy to aggregate local results from the agents, these methods enjoy low communication cost but cannot achieve satisfying statistical efficiency. 
In contrast, \cite{zhu2021least} proposed a weighted least-square approach to aggregate the local estimates, which significantly improves the statistical efficiency. However, this method requires each agent transmitting a high dimensional matrix to the central server, which is unbearable in scenarios where communication costs are high. 
For the case of MSL(multi-shot learning), it takes an iterative strategy to refine the results via multiple rounds of communications between agents and the central server. Notable works  include but not limited to \cite{jordan2018communication},\cite{tan2022communication} and \cite{tu2023byzantine} for the sparse parametric models, as well as \cite{guo2025accelerated} for the multivariate Gaussian mixture models.
It is worth noting that while MSL can achieve optimal statistical efficiency,  it often incurs a heavy communication burden, particularly in high-dimensional settings. Therefore, MSL is not appropriate for scenarios where communication is costly or resource-constrained. 

The above discussion arises an important question: Is there a one-short distributed learning method can achieve satisfactory statistical efficiency while maintaining low communication costs? This has been a critical challenge in distributed learning systems.

To address this problem, we introduce a weighted approach to ensemble the local estimates, in which the weight assigned to each coordinate across different agents is precisely proportional to the inverse of the variance of the local estimates for that agent. Intuitively, a local estimator with smaller variance   would be assigned larger weight, as it is more reliable. 
It is shown that such a weight is optimal in the sense of minimizing the mean squared error (MSE) of the estimates, provided that the local estimators are consistent. This result also implies delivering the entire Hessian matrix is not necessary, but only increasing the communication cost. Overall, the new method mainly sheds light on the following two aspects:
\begin{enumerate}
	\item[(a)] Compared to the existing OSL approaches, it can achieve the same optimal statistical efficiency as the least-square weighted aggregation proposed by \cite{zhu2021least}.
	
	\item[(b)] It achieves remarkably low communication costs, as each agent only needs to transmit two vectors to the server, keeping the communication cost at \(O(p)\). In comparison, the least-square weighted aggregation strategy incurs a communication cost of \(O(p^2)\). This substantial reduction in communication makes our method highly efficient, especially in high-dimensional scenarios where \(p\) is large. 
\end{enumerate}

The contribution of this article is at two-folds: 
Methodologically, we introduce a novel one-shot learning strategy for estimating high-dimensional sparse parameters, which is applicable to a wide range of scenarios, including but not limited to least-square loss, Huber loss, and likelihood-based models. This approach provides a flexible and efficient framework for distributed inference in high-dimensional settings.  
Theoretically, we establish both the error bounds and the asymptotic normality for the parameter estimates. These theoretical guarantees ensure the reliability and safe application of the proposed strategy in distributed learning systems, providing a solid foundation for its practical implementation.  

Finally, for clarity, we introduce some notations. Throughout the paper, we use the bold uppercase letters such as $\bf M$ to represent a matrix, the italic bold uppercase or lowercase letters such as $\bm V$ or $\bm v$ to represent the vectors. The Euclidean norm of a vector $\bm v$ is denoted by $\|\bm v\|_2$. A diagonal matrix consisted of $a_1,\cdots,a_p$ is denoted by $\mathrm{diag}(a_1, a_2, \cdots, a_p)$. We use $\lambda_{\min}({\bf M})$ or $\lambda_{\max}({\bf M})$ to denote the minimum or maximum eigenvalue of $\bf M$. Moreover, let ${\bf M}^{\mathcal A}$ be the sub-matrix of $\bf M$, with the indices of rows and columns belong to $\mathcal A$. Additionally, to clear clarify our method, we introduce a concept of communication cost, defined as the number of elements each agent needs to transmit to the central server. For example, transmitting a $p$-dimensional vector from the agent incurs a communication cost of $p$.

The organization of the paper is as follows. Section 2 gives the details of the methodological development. Section 3 conducts some solid theoretical analyses on our method. Section 4 presents the numerical results to show the effectiveness of the new approach. Section 5 conducts a real data analysis. Section 6 concludes the present article and discusses some pros and cons of new method. 

\section{Methodologies}
	Let $\left\{{\bm X}_i, Y_i \right\}_{i=1}^N$ be a set of identically independently distributed (i.i.d) samples distributed over $K$ agents, where ${\bm X}_i=\left(X_{i}^{(1)}, X_{i}^{(2)}, \ldots, X_{i}^{(p)}\right)^\top$ is a $p$-dimensional covariate and $Y_i$ is the response variable. Here, we consider the case that $p$ is of order $O(n^\nu)$ for some constant $\nu<1$. 
	
	Suppose that the response $Y_i$ associates with the covariate $\bm X_i$ via its linear affine transformation. More specifically, we suppose that the parameter of interest is the minimizer of the following loss function,
	\begin{equation}\label{loss0}
	\boldsymbol{\beta}_\star=\mbox{argmin}_{\boldsymbol{\beta} \in \mathbb{R}^{p}} \sum_{i=1}^N\mathcal{L}\left(Y_i, \boldsymbol{X}_i^\top\bm\beta\right),
	\end{equation}
	where $\mathcal{L}(Y,z)$ is convex and twice-differentiable with respect to $z$, and vector $\boldsymbol{\beta}_\star$ is assumed to be sparse, i.e. the number of nonzero entries in $\boldsymbol{\beta}_\star$ is small compared with $p$. 
	Framework (\ref{loss0}) covers several common losses or models, the following are some typical examples:
	\begin{description}
		\item [(a)] Least square loss: $\mathcal L(y,b)=(y-b)^2$,
		\item [(b)] Huber loss: $\mathcal L(y,b)=h_a(y-b)$, where 
		\begin{equation*}
		h_a(x)=\left\{\begin{array}{ll}
		\frac12x^2, & \mbox{if~}|x|<a,\\
		a|x| - \frac12a^2, & \mbox{otherwise},
		\end{array}\right.
		\end{equation*}
		\item [(c)] Negative log-likelihood: $\mathcal L(y,b)=-\log\{\mathrm{e}^{yb}/(1+\mathrm{e}^b)\}$ for logistic model and $\mathcal L(y,b)=-yb+\mathrm{e}^b$ for Poisson log-linear regression.
	\end{description}
	
	However, directly solving (\ref{loss0}) is difficult because the central server has no access to the entire data set in the distributed learning system, but only the permission to receive some summary statistics from the agents. In such a situation, the distributed learning strategy is a powerful tool to estimate the parameter. 

	For notation simplicity, we denote by $\mathcal I_j$ the index set of the samples belonging to the $j$-th agent, with size $n_j$, satisfying that $\mathcal I_j \cap \mathcal I_k=\emptyset$ for $j\neq k$ and $\cup_{j=1}^K \mathcal I_j=\{1,\dots,N\}$.
	Then, only using the samples in $j$-th agent, and simultaneously taking the sparsity of $\bm\beta_\star$ into consideration, $\bm\beta_\star$ can be estimated as
	\begin{equation}\label{loss_adalasso}
	\widehat{\bm\beta}_{j}=\mathop{\mbox{argmin}}_{\bm\beta} \sum_{i\in\mathcal I_j}\mathcal{L}\left(Y_{i},\boldsymbol{X}_{i}^{\top} \boldsymbol{\beta}\right)+\mathcal P\left(\kappa_j,\bm\beta\right),
	\end{equation}
	where $\mathcal P\left(\kappa,\bm\beta\right)$ is a penalty function and $\kappa_j$ is a tuning parameter for $j$-th agent. In this paper, we employ the adaptive-lasso type penalty on $\bm\beta$, namely, $$\mathcal P(\kappa_j,\bm\beta)=\kappa_j\sum_{d=1}^p\hat\zeta_{j}^{(d)}\left|\beta^{(d)}\right|$$
	where $\hat\zeta_j^{(d)}=|\beta^{(d)}_{\mathrm{Lasso},j}|^{-\alpha}$ is a weight factor, with $\widehat{\beta}_{\mathrm{Lasso},j}^{(d)}$ the lasso estimator of $\beta_\star^{(d)}$. 
	It is known that the Adaptive Lasso penalty has the so called ``Oracle property'' that it can provide consistent estimator. 

	With the obtained estimation $\widehat{\bm\beta}_j$, a commonly employed aggregation method is to use the simple average estimator(SAVE), defined as
	\begin{equation}\label{AV}
	\widehat{\bm\beta}_{\mathrm{SAVE}}=\sum_{j=1}^K \alpha_j\widehat{\bm\beta}_{j},
	\end{equation}
	where $\alpha_j=n_j/N$.
	However, this kind of aggregation overlooks the scale differences between $\widehat{\bm\beta}_j$'s, which might lead to the efficiency loss of the distributed learning. In fact, (\ref{AV}) can be seen as a special case of the following coordinate-wise weighted average estimator, defined as
	\begin{equation}\label{weight-AV}
		\left(\sum_{j=1}^Kw_{j}^{(1)}\widehat{\beta}_j^{(1)},\cdots,\sum_{j=1}^Kw_{j}^{(p)}\widehat{\beta}_j^{(p)}\right)^\top=\sum_{j=1}^K\bm W_j\widehat{\bm\beta}_j,
	\end{equation}
	where $\bm W_j=\mbox{diag}\left(w_{j}^{(1)},\cdots,w_{j}^{(p)}\right)$ for $j=1,\cdots,K$. Obviously, setting $w_j^{(d)}=\alpha_j$ for all $j$ will reduce (\ref{weight-AV}) to (\ref{AV}).
	Our goal is to find a $\bm W_j$ that minimize the variance of $\sum_{j=1}^K\bm W_j\widehat{\bm\beta}_j$, namely,  
	\begin{equation}\label{MSE}
		\mathbb E\left\|\sum_{j=1}^K\bm W_j\widehat{\bm\beta}_j-\bm\beta_\star\right\|_2^2.
	\end{equation}
The following proposition provides a closed form for $\bm W_j$  under some conditions. 	

	\begin{prop}\label{prop1}
	Suppose that $\widehat{\bm\beta}_j$ is a (asymptotically)consistent estimate of $\bm\beta_\star$, with its covariance matrix $n_j^{-1}{\bm\Sigma}_j$, then the solution to (\ref{MSE}) is 
	$${\bm W}_j^o=\mathrm{diag}\left(\frac{n_j\sigma_{j,11}^{-2}}{\sum_{j=1}^K n_j\sigma_{j,11}^{-2}},\cdots,\frac{n_j\sigma_{j,pp}^{-2}}{\sum_{j=1}^K n_j\sigma_{j,pp}^{-2}}\right),$$ 
	where $\sigma_{j,st}^2$ is the $(s,t)$-th element of ${\bm\Sigma}_j$.
	\end{prop} 
	The proof of Proposition \ref{prop1} is deferred in Appendix. Particularly, when the samples across agents are i.i.d from the same distribution, namely, $\sigma^2_{j,dd}=\sigma_{k,dd}^2$, $w_j^{(d)}$ reduces to $n_j/N$. 
	With the optimal weight, we define the weighted average estimator(WAVE) as 
	\begin{equation}\label{weight-AV1}
			\widehat{\bm\beta}_{\mathrm{WAVE}}^o=\sum_{j=1}^K\bm W_j^o\widehat{\bm\beta}_j=\left(\sum_{j=1}^Kn_j\bm\Lambda_j^{-1}\right)^{-1}\left(\sum_{j=1}^Kn_j\bm\Lambda_j^{-1}\widehat{\bm\beta}_j\right),
	\end{equation}
	where $\bm\Lambda_j=\mbox{diag}(\bm\Sigma_j)=\mbox{diag}(\sigma_{j,11}^{2},\cdots,\sigma_{j,pp}^{2})$. 
	By proposition \ref{prop1}, it has that 
	$\mathbb E\|\widehat{\bm\beta}_{\mathrm{WAVE}}^o-\bm\beta_\star\|_2^2\leq E\|\widehat{\bm\beta}_{\mathrm{SAVE}}-\bm\beta_\star\|_2^2.$

The remaining problem is to estimate $\widehat{\bm\Lambda}_j$.  Under the framework (\ref{loss0}), by Taylor expansion, it has that 
\begin{equation}\label{asymptotic_represent}
	\bm\beta_\star-\widehat{\bm\beta}_j=
	\left(\frac{1}{n_j}\sum_{i\in\mathcal I_j}\bm X_i\bm X_i^\top\mathcal L^{\prime\prime}_i(\bm\beta_\star)\right)^{-1}\left(\frac{1}{n_j}\sum_{i\in\mathcal I_j}\bm X_i\mathcal L^\prime_i(\bm\beta_\star)\right) + O_p\left(\left\|\widehat{\bm\beta}_j-\bm\beta_\star\right\|_2^2\right)
\end{equation}
where $\mathcal L_i^\prime(\bm\beta)=\mathcal L^\prime(Y_i,\bm X_i^\top\bm\beta)$ and  $\mathcal L_i^{\prime\prime}(\bm\beta)=\mathcal L^{\prime\prime}(Y_i,\bm X_i^\top\bm\beta)$. Here $\mathcal L^\prime(Y,z)$ is the first-order derivative of $\mathcal L(Y,z)$ respect to $z$, similar definitions goes to $\mathcal L^{\prime\prime}(Y,z)$. 
Neglecting the tail in (\ref{asymptotic_represent}), and by plug-in method, we estimate $\bm\Sigma_j$ as $\widehat{\bm\Sigma}_j=\widehat{\bm\Phi}_j^{-1}\widehat{\bm\Psi}_j\widehat{\bm\Phi}_j^{-1}$
with
\begin{equation}\label{estimate_cov}
	\widehat{\bm\Phi}_j=\frac1{n_j}{\bf X}_j^\top\widehat{\bf L}_j^{\prime\prime}{\bf X}_j\quad\mbox{~and~}\quad\widehat{\bm\Psi}_j=\frac{1}{n_j}{\bf X}_j^\top(\widehat{\bf L}^\prime_j)^2{\bf X}_j,
\end{equation}
where ${\bf X}_j$ is the $n_j\times p$ data matrix, $\widehat{\bf L}_j^\prime$ and $\widehat{\bf L}_j^{\prime\prime}$ are two $n_j\times n_j$ diagonal matrices with elements consisted of $\mathcal L_i^\prime(\widehat{\bm\beta}_j)$ and $\mathcal L_i^{\prime\prime}(\widehat{\bm\beta}_j)$ for $i\in\mathcal I_j$, respectively. 
Finally, with $\widehat{\bm\Lambda}_j=\mathrm{diag}\left(\widehat{\bm\Sigma}_j\right)$, we obtain the estimate of $\bm\beta_\star$  as
\begin{equation}\label{weight-AV2}
		\widehat{\bm\beta}_{\mathrm{WAVE}}=\left(\sum_{j=1}^Kn_j\widehat{\bm\Lambda}_j^{-1}\right)^{-1}\left(\sum_{j=1}^Kn_j\widehat{\bm\Lambda}_j^{-1}\widehat{\bm\beta}_j\right).
	\end{equation}

	Since $\bm\beta_\star$ is assumed to be sparse while $\widehat{\bm\beta}_{\mathrm{WAVE}}$ may not be, we encourage the sparsity of $\widehat{\bm\beta}_{\mathrm{WAVE}}$ by applying an adaptive-lasso type penalty. The final resulting estimator is defined as 
	\begin{equation}\label{ada_aggregate}
		\widetilde{\bm\beta}_{\mathrm{WAVE}}=\mathop{\mbox{argmin}}_{\bm\beta}\frac1{2}\bm\beta^\top\sum_{j=1}^Kn_j\widehat{\bm\Lambda}_{j}^{-1}\bm\beta-\bm\beta^\top\sum_{j=1}^Kn_j\widehat{\bm\Lambda}_{j}^{-1}\widehat{\bm\beta}_{j}+\delta\sum_{d=1}^p\hat\gamma^{(d)}|\beta^{(d)}|,
	\end{equation}
	where $\hat\gamma_d$ can be set as $\left|\hat\beta^{(d)}_{\mathrm{WAVE}}\right|^{-\alpha_0}$ for some $\alpha_0>0$. It can be easily calculated that $\widetilde{\bm\beta}_{\mathrm{WAVE}}$ has a closed expression with its $d$-th coordinate equal to 
	\begin{equation}\label{soft-thresh}
		\widetilde{\beta}_{\mathrm{WAVE}}^{(d)}=\mbox{sign}\left(\widehat{\beta}^{(d)}_{\mathrm{WAVE}}\right)\left(\left|\widehat{\beta}^{(d)}_{\mathrm{WAVE}}\right|-\frac{\delta\hat\gamma^{(d)}}{\sum_{j=1}^Kn_j\hat\sigma_{j,dd}^{-2}}\right)_{+}.
	\end{equation}
	In practice, since $N$ is very large, the tuning on $\delta$ is not necessary, we direct set it equal to $\sqrt{\log p}$, making the thresholding value with order of $O_p(\sqrt{\log p/ N})$, this thresholding value is a little larger than standard order $O_p(\sqrt{p/N})$, see Theorem \ref{global_oracle} in next section. Our numerical experience also shows that this setting can produce satisfactory results.
	
	Overall, the proposed distributed learning algorithm is summarized as following Algorithm \ref{alg:wave}.
	
	\begin{algorithm}[htbp]
		\caption{Algorithm of WAVE.}\label{alg:wave}
		\LinesNumbered
		\KwIn{Datasets $\left\{\bm X_i, Y_i\right\}$ for $i\in\mathcal I_j,j=1,\cdots,K$ distributed over $K$ agents.} 
		\KwOut{The global estimator $\widetilde{\bm\beta}_{\mathrm{WAVE}}$}
		All agents get the local estimator $\widehat{\bm\beta}_j$ using the formula (\ref{loss_adalasso}) and the corresponding covariance matrix $n_j^{-1}\widehat{\bm\Sigma}_j$\;
		Each agent delivers $\widehat{\bm\beta}_j$,  $\mbox{diag}(\widehat{\bm\Sigma}_j)$ and the sample size $n_j$ to the central server\;
		The central server computes the global estimator $\widetilde{\bm\beta}_{\mathrm{WAVE}}$ using formula (\ref{ada_aggregate}). 
	\end{algorithm}
		
	\textbf{\em Comparison with ADLE and WLSE.}   \cite{battey2018distributed} and \cite{lee2017communication} proposed an averaging debiased lasso estimate (ADLE) as
	\begin{equation}\label{simpleAV}
	\widehat{\bm\beta}_{\mathrm{ADLE}}=\sum_{j=1}^K\alpha_j\widehat{\bm\beta}_{\mathrm{DLasso},j},
	\end{equation}
	where $\widehat{\bm\beta}_{\mathrm{DLasso},j}=\widehat{\bm\beta}_{\mathrm{Lasso},j}+{\bm\Theta}{\bf X}_j^\top({\bf y}_j-{\bf X}_j\widehat{\bm\beta}_{\mathrm{Lasso},j})$ is a debiased lasso estimator, $\bm\Theta$ is a matrix that needs to be learned. 
	One can see \cite{lee2017communication} for more details about the debiased lasso method. 
	Compared with our work, the average debiased lasso estimator is communication-efficient but statistically less efficient, since it also cannot account for the heterogeneous variances of different local estimates. Additionally, learning $\bm\Theta$ is computationally intensive, requiring solving $p$ optimization problems, which is unbearable when $p$ is large. 
	Besides, $\widehat{\bm\beta}_{\mathrm{DLasso},j}$ is not sparse, although we can make a hard thresholding strategy on $\widehat{\bm\beta}_{\mathrm{DLasso},j}$. 
	However, it is worth noting separately that the ADLE has a notable feature that it can handle higher dimensional parameter in linear model. 

	Another notable method is also a weighted least-square estimate (WLSE) proposed by \cite{zhu2021least}, defined as
	\begin{equation}\label{LSE_distributed}
	\widehat{\bm\beta}_{\mathrm{WLSE}}=\left(\sum_{j=1}^K n_j\widehat{\bm\Sigma}_{j}^{-1}\right)^{-1}\left(\sum_{j=1}^K n_j\widehat{\bm\Sigma}_{j}^{-1}\widehat{\bm\beta}_{j}\right).
	\end{equation}
	
	Compared with $\widehat{\bm\beta}_{\mathrm{WAVE}}$, WLSE has a significant drawback: it requires each agent transmitting the entire covariance matrix to the server, which is unbearable for large $p$, especially in some scenarios where the transmission resources are limited. However, WAVE requires each agent to transmit only two $p$-dimensional vectors to the server, which significantly reduces the communication costs from $O(p^2)$ to $O(p)$.
	
	From the perspective statistical efficiency, WAVE is comparable with WLSE. To illustrate this, let us consider an ideal scenario where the covariance matrix $n_j^{-1}\bm\Sigma_j$ of $\widehat{\bm\beta}_j$ is known, then   
	$$\widehat{\bm\beta}_{\mathrm{WLSE}}=\left(\sum_{j=1}^Kn_j\bm\Sigma_j^{-1}\right)^{-1}\left(\sum_{j=1}^Kn_j\bm\Sigma_j^{-1}\widehat{\bm\beta}_j\right).$$
	It can be calculated that the variance of $\widehat{\bm\beta}_{\mathrm{WLSE}}$ and $\widehat{\bm\beta}_{\mathrm{WAVE}}$ are, respectively, 
	$$\mathrm{Var}\left(\widehat{\bm\beta}_{\mathrm{WLSE}}\right)=\left(\left(\sum_{j=1}^K\frac{n_j}{\sigma_{j,st}^{2}}\right)^{-1}\right)_{1\leq s,t\leq p}$$
	and
	$$\mathrm{Var}\left(\widehat{\bm\beta}_{\mathrm{WAVE}}\right)=\left(\sum_{j=1}^K\frac{n_j\sigma_{j,st}^2}{\sigma_{j,ss}^2\sigma_{j,tt}^2}\Bigg /\left(\sum_{j=1}^K\frac{n_j}{\sigma^2_{j,st}}\right)^2\right)_{1\leq s,t\leq p},$$
	which implies that $\mathbb E\left(\hat\beta_{\mathrm{WAVE}}^{(d)}-\hat\beta^{(d)}\right)^2=\mathbb E\left(\hat\beta_{\mathrm{WLSE}}^{(d)}-\hat\beta^{(d)}\right)^2$. This result reveals that, theoretically, the WAVE can achieve the same statistical efficiency as WLSE in terms of MSE of their estimates. 
Our simulations also show that the efficiency of WAVE is comparable with WLSE, and moreover in some scenarios, it even surpasses WLSE. The reason may be that while WLSE incorporates the entire covariance matrix, it also introduces more estimation errors.

	\section{Theoretical analysis}
	This section provides some theoretical analyses to give more insight about WAVE. For the convenience of theoretical derivation, we assume that the sample size over all agents is the same, namely $n$. For space saving, here we only present the main theoretical results, the technical details can be found in Appendix.
	Before giving the main results, we first introduce some necessary conditions.
	\begin{description}
		\item [(C1)] $\mathcal L(Y, z)$ is convex and the second-order derivative is differentiable write respect to $z$.
		\item [(C2)] There exists positive constants $b,c,B,C$ such that for all $j=1,\cdots, K$,
		\begin{description}
		\item[(C2a)] $b\leq \lambda_{\min}\left({\bf X}_j^\top{\bf X}_j/n\right)\leq \lambda_{\max}\left({\bf X}_j^\top{\bf X}_j/n\right)\leq B,$
		\item[(C2b)] $c\leq \lambda_{\min}\left({\bf X}_j^\top{\bf L}_j^{\prime\prime}{\bf X}_j/n\right)\leq \lambda_{\max}\left({\bf X}_j^\top{\bf L}_j^{\prime\prime}{\bf X}_j/n\right)\leq C.$
		\end{description}
		\item [(C3)] $p=O(n^\nu)$ with $0\leq \nu <1$.
		\item [(C4)] $\lim_{n \rightarrow \infty} \frac{\kappa_j}{\sqrt{n}}=0, \quad \lim_{n \rightarrow \infty} \frac{\kappa_j}{\sqrt{n}} n^{(\alpha-1) / 2}=\infty$.
		\end{description}
	Condition (C1) makes some requirement on the degrees of local convexity and smoothness of the loss functions, which ensures the unique solution to (\ref{loss_adalasso}). This condition is often assumed in existing literature, see for example \cite{jordan2018communication}, \cite{zhu2021least} and etc.
	Condition (C2) is a mild condition which is commonly assumed in existing literature like  \cite{fan2001variable}, \cite{zou2006adaptive}, \cite{zhu2021least}  and etc. It restricts that both the matrix ${\bf X}_j^\top{\bf L}_j^{\prime\prime}{\bf X}_j/n$ and ${\bf X}_j^\top{\bf X}_j/n$ should be non-singular. This condition can be easily satisfied when the correlation between covariates is regular. Note that (C2b) reduces to (C2a) under the least square loss.
	Condition (C3) restricts that the dimension of covariate can  not exceed the sample size. Condition (C4) is used to establish the asymptotic normality. This condition implies that $\alpha$ should be larger than 1.

	\begin{theorem}\label{bound_local_para}
	Let $\mathcal A=\left\{d:\beta_\star^{(d)} \neq 0\right \}$, then under conditions (C1)-(C4), it has 
	\begin{description}
		\item[(1)] The bound of $\widehat{\bm\beta}_j$ is 
		\begin{eqnarray*}
			\mathbb E\|\widehat{\bm\beta}_j-\bm\beta_\star\|^2_2\leq \frac{8\kappa_j^2\sum_{d=1}^p(\hat\zeta_{j}^{(d)})^2+2npB\mu_j}{n^2c^2},
		\end{eqnarray*} 
		where $\mu_j=\mathbb E\left\{\left(\mathcal L^\prime_i(\bm\beta_\star)\right)^2\right\}$ for $i\in\mathcal I_j$. 
		\item[(2)] $\mathbb P\left(\widehat{\beta}_j^{(d)}=0\right)\rightarrow_p 1$ for $d\in\mathcal A^c$.
		\item[(3)] $\sqrt{n}\left(\widehat{\bm\beta}_j^{\mathcal A}-\bm\beta_\star^{\mathcal A}\right)\sim N\left(\bm 0, \left(\bm\Phi^{\mathcal A}_j\right)^{-1}\bm\Psi^{\mathcal A}_j\left(\bm\Phi^{\mathcal A}_j\right)^{-1}\right).$
		\end{description}
	\end{theorem}
	It is worth mentioning that the derived risk bounds of $\|\widehat{\bm\beta}_j-\bm\beta_\star\|_2$ is of order $O_p(\sqrt{p/n})$, which indicates that $\widehat{\bm\beta}_j$ is a root-($n/p$)-consistent estimator. 
	Theorem \ref{bound_local_para} also tells us that the general loss with adaptive lasso penalty still enjoys the Oracle property. Moreover, Theorem \ref{bound_local_para} also shows that $\widehat{\bm\beta}_j^{\mathcal A}$ is asymptotical normal.
	
	The following theorem presents the asymptotic results for $\widehat{\bm\beta}_{\mathrm{WAVE}}$. Before that we introduce two extra conditions as follows.
	\begin{description}
		\item [(C5)] $K=o(N^\xi)$ with $\xi\leq 1/3$.
		\item [(C6)] $\lim_{N\rightarrow\infty}\frac{\delta}{\sqrt N}=0$ and $\lim_{N\rightarrow\infty}\frac{\delta}{\sqrt N}N^{(\alpha_0-1)/2}=\infty$
	\end{description}
	Condition (C5) implies that the number of agents should not be very large. This condition is stricter than the commonly assumed one that $N=o(\sqrt N)$ in \cite{lee2017communication} and \cite{zhu2021least}. The is because we   that the 
	Condition (C6) is similar to that of (C4).
	
	\begin{theorem}\label{global_oracle}
	Under condition (C1)-(C5), it has that
	\begin{description}
		\item[(1)] The bound of $\widehat{\bm\beta}_{\mathrm{WAVE}}$ is 
		$$\mathbb E\|\widehat{\bm\beta}_{\mathrm{WAVE}}-\bm\beta_\star\|_2^2\leq  \frac{8\kappa_j^2{{\sum_{d=1}^p\left(\zeta_j^{(d)}\right)^2}}K+2{NpB\mu_j}}{N^2c^2}.$$
		\item[(2)] $\mathbb P\left(\widehat{\beta}_{\mathrm{WAVE}}^{(d)}=0\right)\rightarrow_p 1$ for $d\in\mathcal A^c$.
		\item[(3)] If~$\mathbb E\left\{\mathcal{L}^\prime_i(\bm\beta_\star)X_{i}^{(d)}\right\}^4 < \infty$ for $i\in\mathcal I_j, j=1,\cdots,K$,
		\begin{equation*}
			\bm a^\top \left(\widehat{\bm\beta}^{\mathcal A}_{\mathrm{WAVE}}-\bm\beta^{\mathcal A}_\star\right)\sim N\left(\bm 0, \frac1N\bm a^\top \bm V_K^{-1}\bm S_K\bm V_K^{-1} \bm a\right),
		\end{equation*}
		where $ \bm V_K=\frac1K\sum_{j=1}^K\left(\bm\Lambda^{\mathcal A}_j\right)^{-1}$ and $\bm S_K=\frac1K \sum_{j=1}^K\left(\bm\Lambda_j^{\mathcal A}\right)^{-1}\bm\Sigma_j^{\mathcal A}\left(\bm\Lambda_j^{\mathcal A}\right)^{-1}$. 
	\end{description}
	\end{theorem}
	Note that we add a quite stricter condition $\mathbb E\left\{\mathcal{L}_i^\prime(\bm\beta_\star)X_{i}^{(d)}\right\}^4 < \infty$, to guarantee the asymptotic normality of $\widehat{\bm\beta}_{\mathrm{WAVE}}$. This is because we allow the samples from different agents could be with different distributions.
	
	\begin{coro}\label{tilde_beta}
		Let $\hat\gamma^{(d)} =\left|\hat{\beta}_{\mathrm{WAVE}}^{(d)}\right|^{-\alpha_0}$, then under the condition (C6), it has that
		\begin{description}
		\item[(1)] $\widetilde{\beta}_{\mathrm{WAVE}}^{(d)}-\widehat{\beta}_{\mathrm{WAVE}}^{(d)}=O_p\left(1/N\right)$ for $d\in\mathcal A$,
		\item[(2)] $\widetilde{\beta}_{\mathrm{WAVE}}^{(d)}=O_p\left(1/\sqrt N\right)$ for $d\in\mathcal A^c$.
		\end{description} 
	\end{coro}

	\section{Numerical studies}
	We compare the new method WAVE with three commonly used one-shot methods: 
	the simple average method (SAVE) defined in (\ref{AV}), the averaged debiased lasso estimator (ADLE) proposed by \cite{lee2017communication} and \cite{tang2020distributed}, and the distributed weighted least-square estimate WLSE proposed by \cite{zhu2021least}.
	For all methods, the local tuning parameter is set as $\sqrt{\log p/n}$ to guarantee the sparsity of local estimates. In final aggregation step, SAVE directly uses $\sqrt{\log p/N}$ as the threshold value, ADLE follows the methods in \cite{lee2017communication} and \cite{tang2020distributed}, in which \cite{lee2017communication} takes the soft-threshold rule and \cite{tang2020distributed} takes the voting method. WLSE uses the BIC criterion to select the tuning parameter.
	Besides, according to Condition (C4) and (C6), we chose $\alpha=1.5$ and $\alpha_0=1.5$, respectively. 
	
	We employ the mean squared error(MSE) to evaluate the quality of an estimator, which is defined as
	$$\mbox{MSE}(\widehat{\bm\beta})=\frac1T\sum_{t}^T\|\widehat{\bm\beta}(t)-\bm\beta_\star\|_{2}^2,$$
	where $\widehat{\bm\beta}(t)$ is the estimation of $\bm\beta_\star$ in $t$-th simulation. We repeat the simulation 200 times.

\begin{figure}[h!]
\centering
	\includegraphics[width=1\textwidth]{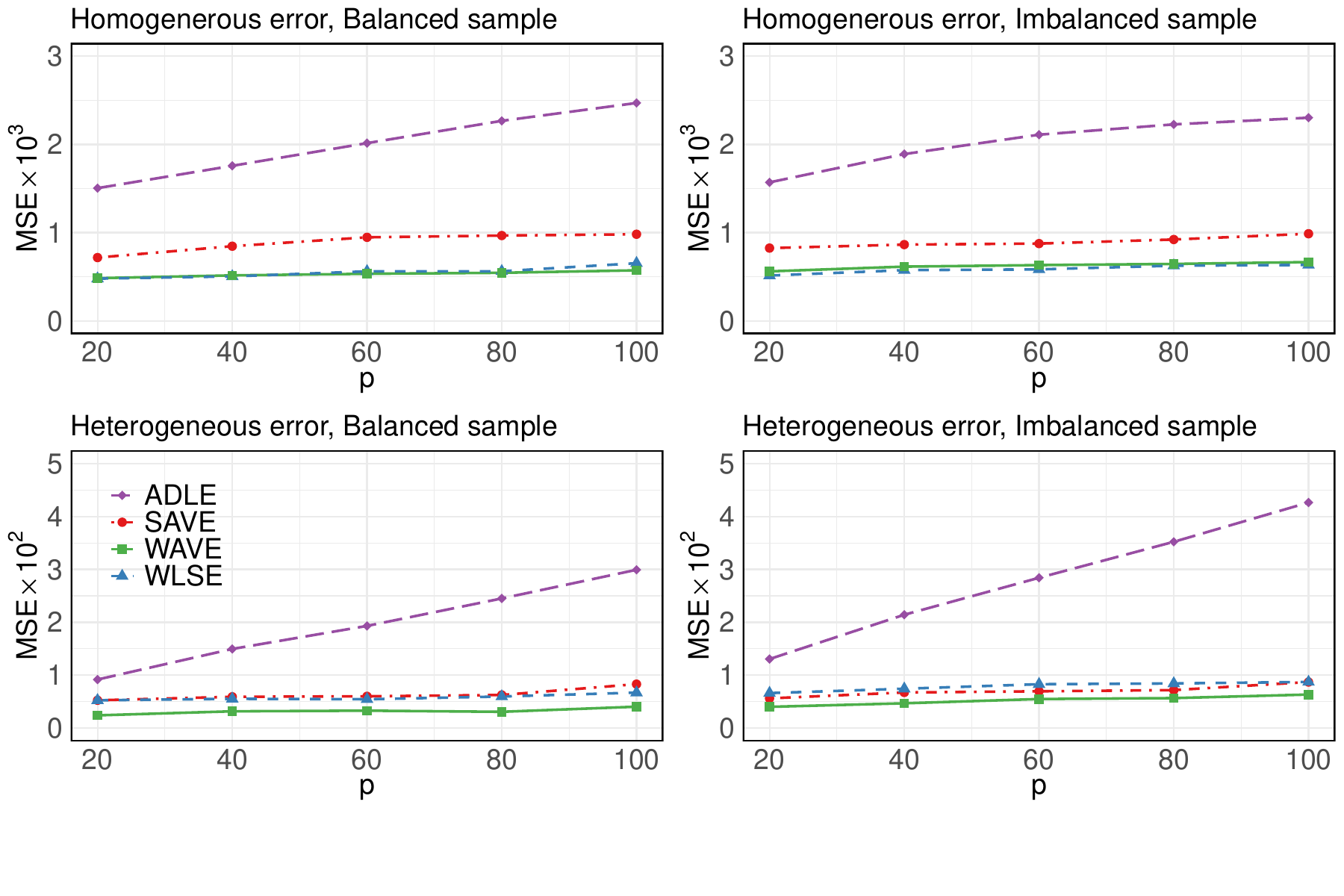} 
	\caption{The Mean Squared Error(MSE) of $\bm\beta$ for different methods across various $p$ in Experiment 1, under the least square loss.}\label{fig:ex1_ls}
\end{figure}

\begin{figure}[h!]
\centering
	\includegraphics[width=1\textwidth]{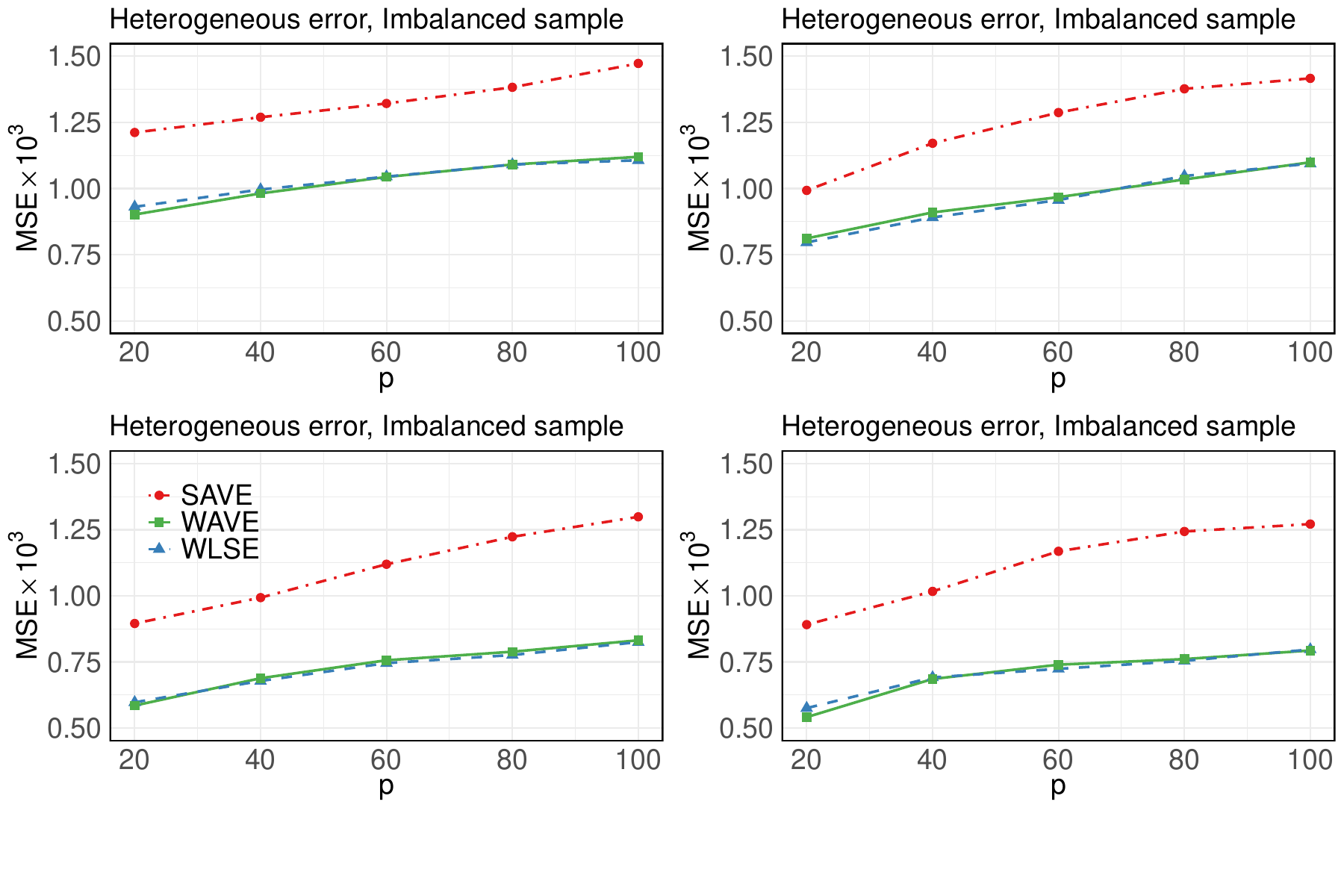} 
	\caption{The MSE of $\bm\beta_\star$ for different methods across various $p$ in Experiment 1, under the huber loss.}\label{fig:ex1_huber}
\end{figure}

	\textbf{Experiment 1}. The samples are generated from the model $Y_i=\bm X_i^\top\bm\beta_\star+\varepsilon_i$, where $\bm X_i$ is from $N(\bm 0, \bm S)$ with $\bm S=(0.5^{|i-j|}), 1\leq i,j\leq p$. The random error $\varepsilon_i$ is from $N(0,s^2_k)$ for $i\in\mathcal I_k$. We consider two different settings for $s_k^2$: (a) homogeneous error variance, $s_k^2=K$ for $k=1,\cdots,K$ and (b) heterogeneous error variance, $s_k^2=k$ for $k=1,\cdots,K$. The design of case (b) makes the data across agents with very different variance.
	We set $N=10,000$, $p\in\{20,40,60,80,100\}$ and the number of agents as $K=N^{1/3}\approx 20$, which is consistent to the condition (C5). The sample size in different agents is considered with two settings: (1) Balanced setting with $n_k=500$ for all $k$ and (2) Imbalanced setting with $n_k=300 + 100 \times \left\lfloor \frac{k-1}{4} \right\rfloor $ for $k = 1, 2, 3, \ldots, 20$. 
	The true parameter is set as $\bm\beta_\star=(3,1.5,0,0,2,\bm 0_{p-5})^\top$. 
	We employ the least-square loss and Huber loss to estimate the parameter $\bm\beta_\star$, where in Huber loss, we fix the parameter $a=1.345$, which is an advised rule in \cite{huber2004robust}. The simulation results under least square loss and Huber loss are presented in Figure \ref{fig:ex1_ls} and \ref{fig:ex1_huber}, respectively, from which the following conclusions can be summarized. 
	\begin{enumerate}
	\item Figure \ref{fig:ex1_ls} reveals that under the least square loss, WAVE has comparable performance to WLSE in homogeneous error settings, while exhibiting slightly superior performance in heterogeneous error settings. Moreover, it can be observed that WAVE consistently outperforms both SAVE and ADLE across all scenarios.
	These observations validates our theoretical analysis that WAVE is able to not only achieve the same statistical efficiency as WLSE but also maintains low communication costs at the level of SAVE. 
	
	\item Figure \ref{fig:ex1_huber} illustrates that under Huber loss, WAVE and WLSE show comparable performance, both significantly outperforming SAVE. This phenomenon further support our argument that it is unnecessary to transmit  the entire covariance matrix of the estimate to the central server. Note that under Huber loss, the ADLE method is undefined.
	
	\item It can be seen from both figures that WAVE shows its superior performance under the heterogeneous settings, it even outperforms WLSE under the least-square loss.    
	\end{enumerate}

\begin{figure}[h!]
	\centering
	\includegraphics[width=1\textwidth]{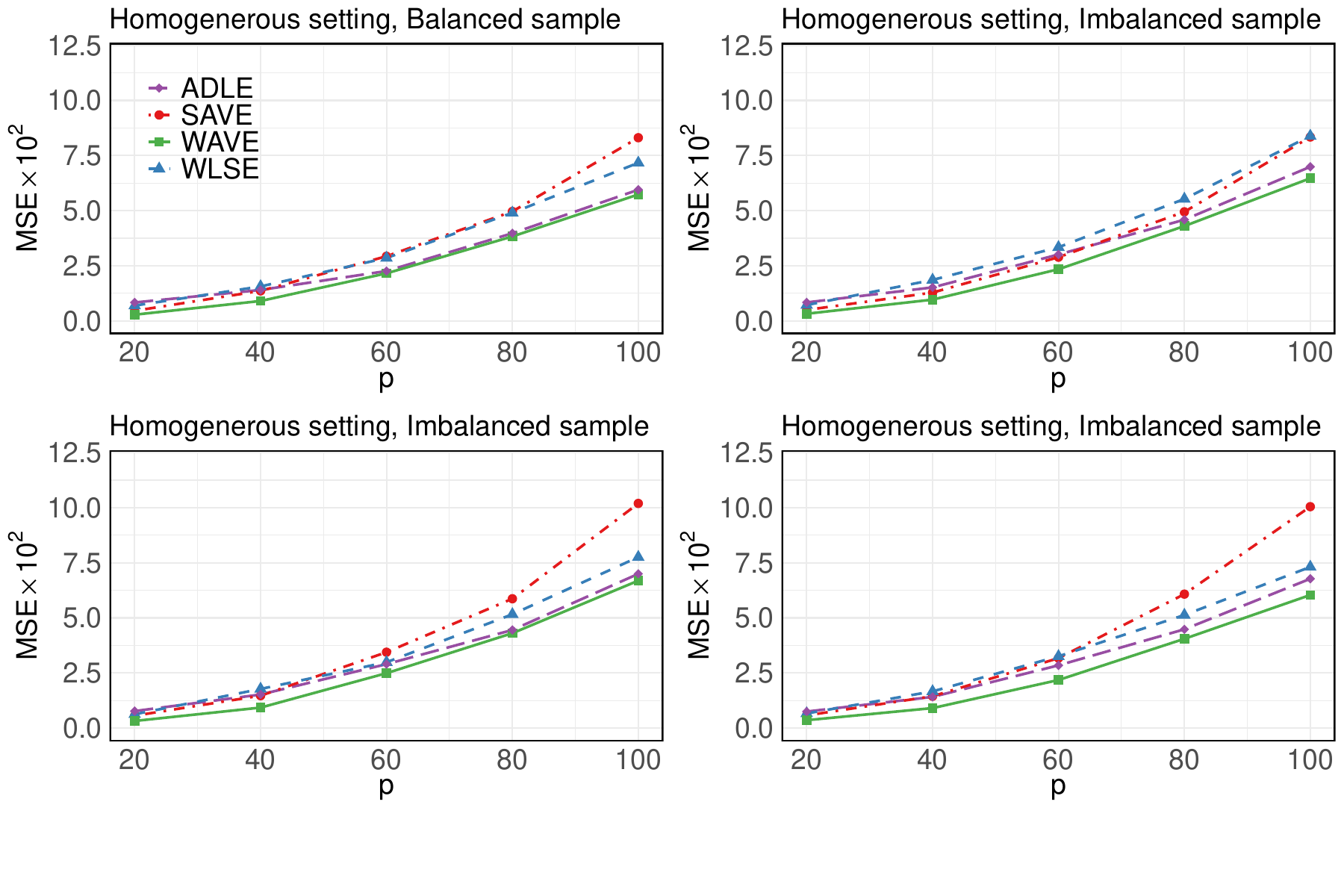} 
	\caption{The MSE of $\bm\beta_\star$ for different methods across various $p$ in Experiment 2.}\label{fig:ex2_logistic}
\end{figure}

	\textbf{Experiment 2}. We further employ a logistic model to check the effectiveness of WAVE. The samples are generated from the model $P(Y_i=1|\bm X_i)=\mathrm{logit}(\bm X_i^\top\bm\beta_\star)$, where $\bm X_i$ is from $N(\bm 0, \bm S_k)$, where $\bm S_k=s_k(0.5^{|i-j|})$. We consider two different setting for $s_k$: (a) Homogeneous setting with $s_k=1$ for $k=1,\cdot,K$; and (b) heterogeneous setting with $s_k=\sqrt{1/4}$ for $k=1,\cdots,4$, $s_k=\sqrt{1/2}$ for $k=5,\cdots,8$, $s_k=1$ for $k=9,\cdots,12$, $s_k=\sqrt{2}$ for $k=13,\cdots,16$, $s_k=\sqrt{4}$ for $k=17,\cdots,20$. Thus, $s_k$ is the source of heterogeneity across different agents.
	We set $N=60,000$, $p\in\{10,20,40,60,80,100\}$.  The number $K$ of agents is still fixed as $20$. The sample size in different agents is considered with two settings: (1) Balanced setting with $n_k=3000$ and (2) Imbalanced setting with $n_k=1800 + 100 \times \left\lfloor \frac{k-1}{4} \right\rfloor $ for $k = 1, 2, 3, \ldots, 20$. The true parameter is still set as $\bm\beta_\star=(3,1.5,0,0,2,\bm 0_{p-5})^\top$. We employ the negative log-likelihood function to estimate the parameter $\bm\beta_\star$. The simulation results are presented in Figure  \ref{fig:ex2_logistic}, from which it can be observed that when the dimension of predictor is not very high, WAVE shows its superior performance over all methods. However, when the dimension $p$ reaches 100, ADLE begins to show its advantage. This is understandable since ADLE utilizes the debiased LASSO for parameter estimation, making it less sensitive to the dimension of predictor. 

	\section{A real-world data application}
	In this section, to assess the performance of our proposed WAVE method in practical large-scale data applications, we study the finite-sample performance of WAVE using a real-world dataset. 
	We employ a countrywide traffic accident dataset, which covers 49 states of the United States. The dataset is continuously collected from February 2016 to March 2023, using several data providers, including multiple APIs that provide streaming traffic event data. These APIs broadcast traffic events captured by a variety of entities, including the state departments of transportation, law enforcement agencies, traffic cameras, and traffic sensors within the road-networks.
	
	This dataset contains 7,728,394 records and each record has 46 attributes, including the severity of the accident, the length of the road extent affected by the accident, the temperature, wind chill, humidity and etc. For more information about his dataset, please visit \url{https://www.kaggle.com/datasets/sobhanmoosavi/us-accidents}.  
	This dataset can be used for numerous applications, such as real-time car accident prediction, studying car accident hotspot locations, casualty analysis, extracting cause and effect rules to predict car accidents and predicting the severity of the accident. Among these, we select the accident severity as the prediction target in our study, as it is a critical factor in evaluating the impact of traffic accidents. 
	This focus aligns the goal of enhancing traffic safety and emergency response strategies \citep{moosavi2019countrywide}. 
		
	We build a logistic regression model to describe the relationship between the predictors and the response variable, namely, the severity of the accident. The original dataset categorized accident severity into four levels, ranging from 1 to 4. To simplify the analysis, we transformed this four-level variable into a binary one, with 0 representing minor accidents and 1 representing severe accidents.
	
\begin{figure}[h!]
\centering
	\includegraphics[width=0.9\textwidth]{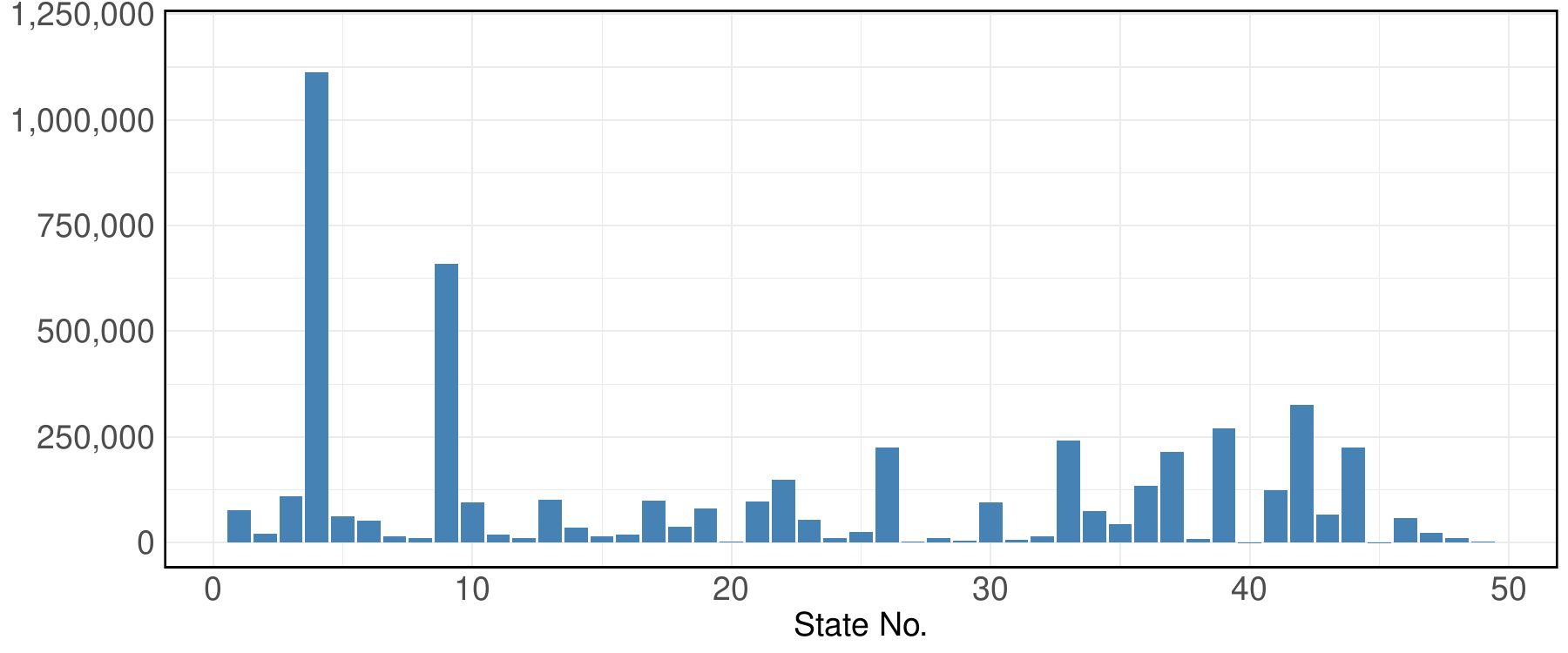} 
	\caption{The number of records of the traffic accidents in 49 states.}\label{fig:sample_size}
\end{figure}

	Directly modeling this dataset presents several challenges. First, the dataset's size is too large to load into the memory of a standard PC, making traditional analysis methods infeasible. Second, as \cite{moosavi2019accident} pointed out, the accident data is inherently heterogeneous, meaning it should not be treated as a single homogeneous population. 
	To adapt the heterogeneity in the dataset, we divided the large scale dataset into 49 subsets based on the state. Given that the severity of an accident may be influenced by the geographical location of a state, we assumed that each subset could exhibit homogeneous error variance within its respective region. 
	Moreover, Figure \ref{fig:sample_size} also displays the number of records across different states. It can be seen that the sample sizes vary significantly from state to state, implying that the variances of the estimates across states are also very different. Consequently, taking the simple average strategy to aggregate the local results may not be an effective method.

Since the dataset is up to March 31, 2023, we select the records of March 31, 2023---the most recent day in March---as the testing samples, totally having 1904 records. The records prior to this date are divided into 49 subsets as the training samples according to the state. 
This strategy not only makes the current statistical methods applicable in handling this kind of large scale dataset, but also offers considerable flexibility as each state is only required to transmit two representative 46-dimensional vectors to the designated devices. 
We apply the WAVE along with the three competitors respectively to the dataset to infer the parameter in the logistic model. After that, we use the testing samples to check the performance of the model under different methods, the corresponding misclassification error rates on testing samples are presented in Table \ref{tab:data_anal}, from which it can be seen that  the WAVE and WLSE exhibit comparable performance, with both them superior than the SAVE and ADLE methods. However, WAVE achieves this with a significantly lower communication cost compared to WLSE, which requiring transmit a 46-dimensional matrix (with a communication cost of 1058). This result highlights the effectiveness of WAVE.

\begin{table}
\centering
\caption{The misclassification error rates for different methods on the testing samples.}\label{tab:data_anal}
\begin{tabular*}{\hsize}{@{}@{\extracolsep{\fill}}rccl@{}}
\toprule 
	WAVE &WLSE & SAVE & ADLE \\
	0.9778&0.9780&0.9685&0.9552\\
\bottomrule
\end{tabular*}
\end{table}

	\section{Conclusions remarks}
	
	This paper proposes a weighted average estimator for high dimensional parameter, which is applicable for the distributed learning environment and the large-scale dataset.
		
	The means of our findings can be summarized in three key aspects. Methodologically, we introduce a novel one-shot learning strategy to estimate the high-dimensional sparse parameters, which is effective for the parameter model under least-square loss, Huber loss, and likelihood function, offering a flexible framework for parallel inference in high-dimensional settings. 
	Theoretically, we derive the error bounds for the mean squared error of the estimator, which achieves the standard rate of $O(p/n)$. Additionally, we establish the asymptotic normality of the parameter estimates. These theoretical results ensure the reliability of the proposed estimator, providing a solid foundation for its practical implementation.  
	Practically, the proposed method is well-suited for the distributed computing environments and large-scale datasets. Moreover, it can be implemented very easily by the practitioners across a wide range of application fields. 
	
	The advantages of WAVE are evident. On the one hand, among the one-shot distributed learning approaches, WAVE can achieve the optimal statistical efficiency (with minimal mean squared error) as the least-square weighted aggregation method such as \cite{lin2011aggregated} and \cite{zhu2021least}. It can also observed from Proposition \ref{prop1} and Theorem \ref{global_oracle} that the bound of MSE achieves the standard rate $O(p/N)$. The simulations also reveals that WAVE behaves comparable the state-of-art method WLSE, sometimes even outperforms this method. On the one hand, WAVE achieves remarkably low communication costs, of which, its communication cost is significantly reduced from the order \(O(p^2)\) to \(O(p)\). This substantial reduction in communication makes our method highly efficient, especially in high-dimensional scenarios where \(p\) is large.
	
	However, WAVE also has at least two limitations. First of all, based on our theoretical analysis, it imposes a strict constraint on the number of agents, specifically requiring $K=o(N^\xi)$ for $\xi\leq 1/3$. This restriction limits its application in some distributed systems with a large number of nodes. Second, the dimension of predictors also cannot exceed the sample size in each agent, otherwise, the consistency of the parameter estimation cannot be guaranteed. These drawbacks highlight potential challenges in some distributed settings with ultrahigh dimensional predictors or lots of agents. 
	
	In our framework, a possible approach to addressing the issues of ultrahigh dimensionality is to take a ``screening+selection'' strategy, see for example \cite{fan2008sure}. This two-stage approach first applies the Sure Independence Screening (SIS) strategy to rapidly reduce the ultrahigh dimensionality to a moderate scale, then employs some sophisticated approaches for further inference. We can use a similar strategy. First, each agent applies the screening method such as SIS \cite{fan2008sure} and DCSIS \cite{li2012feature} to reduce the dimensionality to $d_n$ and then employs the adaptive lasso to get the local parameter estimates. Subsequently, each agent transmits the active set $\widehat{\mathcal A}_j$, the parameter estimation $\widehat{\bm\beta}_{j}^{\mathcal A_j}$ and the estimated covariance  $\hat{\sigma}_{j, dd}$ for $j=1,\cdots,K$ to the central server. Finally, the central server aggregates the local results to get the final estimator as
	\begin{equation*}\label{soft-thresh-ultrahigh}
		\widetilde{\beta}_{\mathrm{WAVE}}^{(d)}=\mbox{sign}\left(\widehat{\beta}^{(d)}_{\mathrm{WAVE}}\right)\left(\left|\widehat{\beta}^{(d)}_{\mathrm{WAVE}}\right|-\frac{\delta\hat\gamma^{(d)}}{\sum_{j=1}^Kn_j\hat\sigma_{j,dd}^{-2}}\right)_{+}.
	\end{equation*}
	where in $\widehat{\beta}_{\mathrm{WAVE}}^{(d)}$,  we set $\widehat{\beta}^{(d)}_j$ equal to zero if $d\in\widehat{\mathcal A}_j^c$ and  $\hat{\sigma}_{j,dd}=1$ for $d\in\widehat{\mathcal A}_j^c$.
	
	The second restriction on the number of agents arises from the estimation of the variance. This limitation can be addressed through a sample splitting strategy. Specifically, half of the samples in agents is allocated for estimating the parameter itself, while the other half are used to estimate the variance of the parameter estimation. This approach effectively decouples the tasks of parameter estimation and variance estimation, thereby relaxing the constraint on the number of agents.
		
	Actually, WAVE can be extended to more general models like semi-parametric models, quantile regression models and etc., in which the parameter of interest can be aggregated via the formula like (\ref{ada_aggregate}). The difference lies at that the diagonal matrix should be customized according to specific model, relevant research can be explored in future study.

	\section*{CRediT authorship contribution statement}
	
	\textbf{Jun Lu}: Conceptualization, Writing-original draft, Methodology, Investigation, Formal analysis, Funding acquisition. 
	\textbf{Xiaoyu Ma}: Writing-review \& editing, Methodology, Software, Funding acquisition. 
	\textbf{Mengyao Li}: Data analysis,  Writing-review \& editing.
	\textbf{Chenping Hou}: Methodology, Supervision, Funding acquisition.
	
	\section*{Acknowledgements}
	We would like to acknowledge support for these projects from Key NSF of China(No. 62136005), the NSF of China(No.61922087, No.12001486), and Scientific Research Program Funds of NUDT (No. 22-ZZCX-016, No. ZK21-20).
	Constructive comments from anonymous  reviewers have helped to improve the manuscript and these are gratefully acknowledged. Special thanks are also given to editors  for their meticulous handling of the manuscript and dedication throughout the review process, which has significantly enhanced the readability and overall quality of this work. 
	
		\section*{Appendix: Proof of main theorems}
	\label{app:theorem}
	
	Before the proof, we introduce some necessary notations. 
	Let $\bm L_j^\prime=(\mathcal L^\prime_i(\bm\beta_\star):i\in\mathcal I_j)$ and $\bm L_j^{\prime\prime}=(\mathcal L^{\prime\prime}_i(\bm\beta_\star):i\in\mathcal I_j)$ be the vectors, ${\bf L}_j^{\prime}=\mbox{diag}(\bm L_j^{\prime})$ and ${\bf L}_j^{\prime\prime}=\mbox{diag}(\bm L_j^{\prime\prime})$ be the diagonal matrices with diagonal elements consisted of the vectors $\bm L_j^\prime$ and $\bm L_j^{\prime\prime}$, respectively. 
	Correspondingly, we define $\widehat{\bf L}_j^{\prime}=\mathrm{diag}(\widehat{\bm L}_j^{\prime})$ and $\widehat{\bf L}_j^{\prime\prime}=\mathrm{diag}(\widehat{\bm L}_j^{\prime\prime})$ with $\widehat{\bm L}_j^{\prime}=(\mathcal L_i^\prime(\widehat{\bm\beta}_j): i\in\mathcal I_j)$ and $\widehat{\bm L}_j^{\prime\prime}=(\mathcal L_i^{\prime\prime}(\widehat{\bm\beta}_j): i\in\mathcal I_j)$, respectively.

	\textbf{Proof of Proposition \ref{prop1}}. We can write
	\begin{eqnarray*}
		\mathbb E\|\widehat{\bm\beta}_{\mathrm{WAVE}}-\bm\beta_\star\|_2^2=\sum_{d=1}^p \mathbb E\left( \sum_{j=1}^K w_j^{(d)}\widehat{\beta}_j^{(d)}-\beta_\star^{(d)} \right)^2.
	\end{eqnarray*}
	Thus, it is sufficient to deal with the item $\mathbb E\left( \sum_{j=1}^Kw_j^{(d)}\widehat{\beta}_j^{(d)}-\beta_\star^{(d)} \right)^2$, in which the optimal weight $w_j^{(d)}$ is defined as
	\begin{eqnarray*}
		w_j^{(d)}&=&\mathop{\mbox{argmin}}_{w_j^{(d)}}\mathbb E\left( \sum_{j=1}^Kw_j^{(d)}\widehat{\beta}_j^{(d)}-\beta_\star^{(d)} \right)^2 + \lambda\left(\sum_{j=1}^Kw_j^{(d)}-1\right)\\
		&=& \mathop{\mbox{argmin}}_{w_j^{(d)}}\sum_{j=1}^K\left(w_j^{(d)}\right)^2\sigma^2_{j,dd} +  \lambda\left(\sum_{j=1}^Kw_j^{(d)}-1\right),
	\end{eqnarray*}
	where the second equality holds because we assume that $\mathbb E\widehat{\beta}_j^{(d)}=\beta_\star^{(d)}$. 
	Consequently, the solution of $w_j^{(d)}$ is 
	$$w_j^{(d)}=\frac{\sigma^{-2}_{j,dd}}{\sum_{j=1}^K\sigma^{-2}_{j,dd}}.$$

	\textbf{Proof of Theorem \ref{bound_local_para}}. We prove the results in this theorem in three steps. 
	
	\textit{Proof of result (1).} 
	Let $\widehat{\bm\beta}_j=(\widehat{\beta}_{j}^{(1)},\cdots,\widehat{\beta}_{j}^{(p)})^\top$ be the regularized estimate on $j$-th agent,  and $\breve{\bm\beta}_j=(\breve{\beta}_{j}^{(1)},\cdots,\breve{\beta}_{j}^{(p)})^\top$ be the unregularized estimate, namely, 
	$$\breve{\bm\beta}_j=\mathop{\mathrm{argmin}}\sum_{i\in\mathcal I_j}\mathcal L(Y_i,\bm X_i^\top\bm\beta).$$
	Since $\widehat{\bm\beta}_j$ is the solution to (\ref{loss_adalasso}), we get that
	$$\sum_{i\in\mathcal I_j}\mathcal L(\widehat{\bm\beta}_j)+\kappa_j\sum_{d=1}^p\zeta_{j}^{(d)}|\widehat\beta_{j}^{(d)}| \leq 
	\sum_{i\in\mathcal I_j}\mathcal L(\breve{\bm\beta}_j) +\kappa_j\sum_{d=1}^p\zeta_{j}^{(d)}|\breve\beta_{j}^{(d)}|,$$
	which results in that
	\begin{equation}\label{bound1}
		\kappa_j \sqrt{\sum_{d=1}^p\left(\zeta_{j}^{(d)}\right)^2} \left\|\widehat{\bm\beta}_j-\breve{\bm\beta}_j\right\|_2
		\geq \kappa_j\sum_{d=1}^p\zeta_{j}^{(d)}\left(|\breve\beta_{j}^{(d)}|-|\widehat\beta_{j}^{(d)}|\right)
		\geq \sum_{i\in\mathcal I_j}\left\{\mathcal L_{i}(\widehat{\bm\beta}_j)-\mathcal L_{i}(\breve{\bm\beta}_j)\right\}.
	\end{equation}
	By Taylor expansion, it has 
	\begin{eqnarray*}
		\sum_{i\in\mathcal I_j}\left\{\mathcal L_{i}(\widehat{\bm\beta}_j)-\mathcal L_{i}(\breve{\bm\beta}_j)\right\}
		=\frac12(\widehat{\bm\beta}_j-\breve{\bm\beta}_j)^\top\left(\sum_{i\in\mathcal I_j}\bm X_i\bm X_i^\top \mathcal L_{i}(\breve{\bm\beta})\right)(\widehat{\bm\beta}_j-\breve{\bm\beta}_j)+o_p(n\|\widehat{\bm\beta}_j-\breve{\bm\beta}_j\|_2^2)
	\end{eqnarray*}
	Additionally, again by Taylor expansion, we have
	$$
	\begin{aligned}
		&\sum_{i\in\mathcal I_j}\bm X_i\bm X_i^\top \mathcal L_{i}(\breve{\bm\beta})
		=\sum_{i\in\mathcal I_j}\bm X_i\bm X_i^\top\left(\mathcal L_{i}({\bm\beta}_\star)+\mathcal L^\prime_{i}({\bm\beta}_\star)\bm X_i^\top(\breve{\bm\beta}_j-\bm\beta_\star)+o_p(\|\breve{\bm\beta}_j-\bm\beta_\star\|_2) \right)\\
		=& \sum_{i\in\mathcal I_j}\bm X_i\left(\mathcal L_{i}({\bm\beta}_\star)+\mathcal L^\prime_{i}({\bm\beta}_\star)\bm X_i^\top(\breve{\bm\beta}_j-\bm\beta_\star)\right)\bm X_i^\top+o_p(n\|\breve{\bm\beta}_j-\bm\beta_\star\|_2) \\
		=& \sum_{i\in\mathcal I_j}\bm X_i\left(\mathcal L_{i}({\bm\beta}_\star)+O_p(\|\breve{\bm\beta}_j-\bm\beta_\star\|_2)\right)\bm X_i^\top+o_p(n\|\breve{\bm\beta}_j-\bm\beta_\star\|_2)\\
		=&{\bf X}_j{\bf L}_j^{\prime\prime}{\bf X}_j^\top +  O_p(n\|\breve{\bm\beta}_j-\bm\beta_\star\|_2)+o_p(n\|\breve{\bm\beta}_j-\bm\beta_\star\|_2).
	\end{aligned}
	$$
	Therefore,  
	\begin{equation}\label{bound2}
		\begin{aligned}
			&\sum_{i\in\mathcal I_j}\left\{\mathcal L_{i}(\widehat{\bm\beta}_j)-\mathcal L_{i}(\breve{\bm\beta}_j)\right\}\\
			=&\frac12(\widehat{\bm\beta}_j-\breve{\bm\beta}_j)^\top\left({\bf X}_j{\bf L}_j^{\prime\prime}{\bf X}_j\right)(\widehat{\bm\beta}_j-\breve{\bm\beta}_j)+O_p(n\|\widehat{\bm\beta}_j-\breve{\bm\beta}_j\|_2^2\|\breve{\bm\beta}_j-\bm\beta_\star\|_2)\\
			\geq & \frac12\lambda_{\min}({\bf X}_j{\bf L}_j^{\prime\prime}{\bf X}_j)\|\widehat{\bm\beta}_j-\breve{\bm\beta}_j\|_2^2+o_p(n\|\widehat{\bm\beta}_j-\breve{\bm\beta}_j\|_2^2),
		\end{aligned}
	\end{equation}
	where the last inequality holds because of condition (C2) and the conclusion $\|\breve{\bm\beta}_j-\bm\beta_\star\|_2=O_p(\sqrt{p/n})$, which can be easily proved by the central limited theorem. 
	
	Combing (\ref{bound1}) and (\ref{bound2}) immediately leads to
	\begin{equation}\label{bound3}
		\left\|\widehat{\bm\beta}_j-\breve{\bm\beta}_j\right\|_2\leq \frac{2\kappa_j \sqrt{\sum_{d=1}^p\left(\hat\zeta_j^{(d)}\right)^2}}{\lambda_{\min}({\bf X}^\top_j{\bf L}_j^{\prime\prime}{\bf X}_j)} + o_p(\|\widehat{\bm\beta}_j-\breve{\bm\beta}_j\|_2).
	\end{equation} 
	On the other hand, we have
	\begin{eqnarray*}
		&&\frac1n\sum_{i\in\mathcal I_j}\bm X_i\mathcal L_{i}^\prime(\breve{\bm\beta}_j)-\frac1n\sum_{i\in\mathcal I_j}\bm X_i\mathcal L^\prime_{i}(\bm\beta_\star)=\frac1n{\bf X}_i{\bf L}_j^{\prime\prime}{\bf X}_j^\top(\breve{\bm\beta}_j-\bm\beta_\star)+o_p(\|\bm\beta_\star-\breve{\bm\beta}_j\|_2),
	\end{eqnarray*}
	which implies that
	\begin{eqnarray*}
		\bm\beta_\star-\breve{\bm\beta}_j=\left({\bf X}_j^\top{\bf L}_j^{\prime\prime}{\bf X}_j\right)^{-1}\left({\bf X}_j^\top{\bm{L}_j^\prime}\right)+o_p(\|\bm\beta_\star-\breve{\bm\beta}_j\|_2).
	\end{eqnarray*}
	For the first item in above equation, it has that
	\begin{eqnarray*}
		&&\mathbb E^2\left\|\left({\bf X}_j^\top{\bf L}_j^{\prime\prime}{\bf X}_j\right)^{-1}\left({\bf X}_j^\top{\bm{L}^\prime}_j\right)\right\|_2
		\leq  \mathbb E\left\|\left({\bf X}_j^\top{\bf L}_j^{\prime\prime}{\bf X}_j\right)^{-1}\left({\bf X}_j^\top{\bm{L}^\prime_j}\right)\right\|_2^2  \\
		&\leq & = \frac{\mathbb E ({\bm{L}^\prime_j}^\top{\bf X}_j{\bf X}_j^\top{\bm{L}^\prime_j})}{\lambda_{\min}^{2}({\bf X}_j^\top{\bf L}_j^{\prime\prime}{\bf X}_j)} 
		= \frac{\mu_j\mbox{tr}({\bf X}_j^\top{\bf X}_j)}{\lambda_{\min}^{2}({\bf X}_j^\top{\bf L}_j^{\prime\prime}{\bf X}_j)} 
		\leq  \frac{p\mu_j\lambda_{\max}({\bf X}_j^\top{\bf X}_j)}{\lambda_{\min}^{2}({\bf X}_j^\top{\bf L}_j^{\prime\prime}{\bf X}_j)} \leq \frac{npB\mu_j}{\lambda_{\min}^{2}({\bf X}_j^\top{\bf L}_j^{\prime\prime}{\bf X}_j)}
	\end{eqnarray*}
	where $\mu_j=\mathbb E\left(\mathcal L_i^\prime(\bm\beta_\star)\right)^2$ for $i\in\mathcal I_j$.
	Consequently,  we have
	\begin{equation}\label{bound4}
		\mathbb E\|\bm\beta_\star-\breve{\bm\beta}_j\|_2
		\leq \frac{\sqrt{npB\mu_j}}{\lambda_{\min}({\bf X}_j^\top{\bf L}_j^{\prime\prime}{\bf X}_j)}+o_p\left(\|\bm\beta_\star-\breve{\bm\beta}_j\|_2\right).
	\end{equation}	
	The inequality (\ref{bound3}) and (\ref{bound4}) yields that 
	$$
	\begin{aligned}
		& \mathbb E\|\widehat{\bm\beta}_j-\bm\beta_\star\|_2\leq \mathbb E\|\widehat{\bm\beta}_j-\breve{\bm\beta}_j\|_2+\mathbb E\|\breve{\bm\beta}_j-\bm\beta_\star\|_2\\
		\leq &\frac{2\kappa_j\sqrt{{\sum_{d=1}^p\left(\zeta_j^{(d)}\right)^2}}+\sqrt{npB\mu_j}}{\lambda_{\min}({\bf X}_j{\bf L}_j^{\prime\prime}{\bf X}_j)}+o_p\left(\|\widehat{\bm\beta}_j-\breve{\bm\beta}_j\|_2+\|\bm\beta_\star-\breve{\bm\beta}_j\|_2\right)\\
		\leq & \frac{2\kappa_j\sqrt{{\sum_{d=1}^p\left(\zeta_j^{(d)}\right)^2}}+\sqrt{npB\mu_j}}{nc}+\mathbb E\left(\|\widehat{\bm\beta}_j-\bm\beta_\star\|_2\right)+o_p(\sqrt{p/n})
	\end{aligned}
	$$
	Above arguments also indicates that $\|\bm\beta_\star-\widehat{\bm\beta}_j\|_2=O_p(\sqrt{p/n})$.
	It can also be proved that 
	$$
	\begin{aligned}
		& \mathbb E\|\widehat{\bm\beta}_j-\bm\beta_\star\|_2^2 \leq  
		2\mathbb E\|\widehat{\bm\beta}_j-\breve{\bm\beta}_j\|_2^2+2\mathbb E\|\breve{\bm\beta}_j-\bm\beta_\star\|_2^2 \\
		\leq & \frac{8\kappa_j^2{{\sum_{d=1}^p\left(\zeta_j^{(d)}\right)^2}}+2{npB\mu_j}}{n^2c^2}+o_p(p/{n}).
	\end{aligned}
	$$
	
	\
	
	\textit{Proof of result (2).}
	Let $\eta=\min _{d \in \mathcal{A}}\left(\left|\beta_*^{(d)}\right|\right)$ and $\hat{\eta}_j=\min_{d \in \mathcal{A}}\left(\left|\hat{\beta}_{Lasso,j}^{(d)}\right|\right)$. Then, by taking similar argument in (\ref{bound3}), we have  
	\begin{equation}\label{bound_subset}
		\left\|\widehat{\bm\beta}_j^{\mathcal A}-\breve{\bm\beta}_j^{\mathcal A}\right\|_2\leq \frac{2\kappa_j\hat\eta_j^{-\alpha}\sqrt{p}}{nc},
	\end{equation} 
	which implies that 
	$$\min_{d\in\mathcal A}|\hat\beta^{(d)}_j|>\min_{d\in\mathcal A}|\breve\beta^{(d)}_j|-\frac{2\kappa_j\hat\eta_j^{-\alpha}\sqrt{p}}{nc}>\min_{d\in\mathcal A}|\beta^{(d)}_j|-\left\|\breve{\bm\beta}_j^{\mathcal A}-{\bm\beta}_\star^{\mathcal A}\right\|_2-\frac{2\kappa_j\hat\eta_j^{-\alpha}\sqrt{p}}{nc},$$
	where the last inequality holds because $\min_{d\in\mathcal A}|\breve\beta^{(d)}_j|>\min_{d\in\mathcal A}|\beta^{(d)}_\star|-\left\|\breve{\bm\beta}_j^{\mathcal A}-{\bm\beta}_\star^{\mathcal A}\right\|_2$.
	It is proved that
	$$\mathbb E\left\|\breve{\bm\beta}_j^{\mathcal A}-{\bm\beta}_j^{\mathcal A}\right\|_2\leq \frac{Bpn\mu_j}{n^2c^2}=O_p(p/n),$$
	and $\frac{2\sqrt{p} \hat{\eta}_j^{-\alpha}}{nc}=O\left(\frac{1}{\sqrt{n}}\right)\left(\frac{\kappa_j \sqrt{p}}{\sqrt{n}} \eta^{-\alpha}\right)\left(\frac{\hat{\eta}_j}{\eta}\right)^{-\alpha}=o_p(1/\sqrt n)$ due to the fact 
	\begin{equation*}
		\begin{aligned}
			& \mathbb E\left(\frac{\hat{\eta}_j}{\eta}\right)^2  \leq 2+\frac{2}{\eta^2} E\left((\hat{\eta}_j-\eta)^2\right)  \leq 2+\frac{2}{\eta^2} E\left(\left\|\widehat{\boldsymbol{\beta}}_j-\boldsymbol{\beta}_\star\right\|_2^2\right) 
			\leq  2+\frac{2}{\eta^2}  \frac{8\kappa_j^2{p\eta^{-2\alpha}}+2{npB\mu_j}}{n^2c^2}.
		\end{aligned}
	\end{equation*}
	Hence, 
	\begin{equation}\label{feature_consist}
		\min_{d\in\mathcal A}|\hat\beta^{(d)}_j|>\eta-O_p(p/n)-o_p(1/\sqrt n).
	\end{equation}

	\
	
	\textit{Proof of result (3).} Define
	\begin{equation}\label{def_Q}
		Q_n(\bm\beta)=\sum_{i=\mathcal I_j}\mathcal L_i(\bm\beta)+\kappa_j\sum_{d=1}^p\hat\zeta_j^{(d)}|\beta^{(d)}|.
	\end{equation}
	and 
	\begin{eqnarray*}
		\Psi_n(\bm u)=\sum_{i=\mathcal I_j}\mathcal L_i\left(\bm\beta_\star+\frac{\bm u}{\sqrt{n}}\right)+\kappa_j\sum_{d=1}^p\hat\zeta_j^{(d)}\left|\beta_\star^{(d)}+\frac{u^{(d)}}{\sqrt n}\right|.
	\end{eqnarray*}
	Let $\hat{\bm u}=\mathop{\mathrm{argmin}}\Psi_n(\bm u)$, then $\widehat{\bm\beta}_j=\bm\beta_\star+\frac{\hat{\bm u}}{\sqrt n}$. 
	Let $\Psi_n(\bm u)-\Psi_n(\bm  0)=V_n(\bm u)$, where 
	\begin{eqnarray*}
		V_n(\bm u)&=&\frac{\bm u}{\sqrt n}\sum_{i\in\mathcal I_j}\bm X_i\mathcal L_i^\prime(\bm\beta_\star)+\frac12\bm u^\top\left(\frac1n\sum_{i\in\mathcal I_j}\bm X_i\bm X_i^\top\mathcal L_i^{\prime\prime}(\bm\beta_\star)\right)\bm u + o_p(\|\bm u\|_2^2/\sqrt n)\\
		&&+\kappa_j\sum_{d=1}^p\hat\zeta_j^{(d)}\left(\left|\beta_\star^{(d)}+ \frac{u^{(d)}}{\sqrt n}\right|-|\beta_\star^{(d)}|\right).
	\end{eqnarray*}
	Then $\hat{\bm u}$ is also the minimizer of $V_n(\bm u)$. We  discuss the limiting behavior for the third item in $V_n(\bm u)$. 
	\begin{itemize}
		\item If $\beta_\star^{(d)}\neq 0$, then $\left|\beta_\star^{(d)}+ u^{(d)}/\sqrt n\right|-|\beta_\star^{(d)}|\rightarrow \mbox{sign}(\beta_\star^{(d)})u^{(d)}/\sqrt n$ and $\hat\zeta_j^{(d)}\rightarrow_p |\beta_\star^{(d)}|^{-\alpha}$, since $\lim_{n\rightarrow \infty}\kappa_j/\sqrt n =0$, by the Slustsky's theorem, it has that
		$$\kappa_j\hat\zeta_j^{(d)}\left(\left|\beta_\star^{(d)}+ u^{(d)}/\sqrt n\right|-|\beta_\star^{(d)}|\right)\rightarrow_p 0.$$
		\item If $\beta_\star^{(d)}= 0$, then $\sqrt n(|\beta_\star^{(d)}+ u^{(d)}/\sqrt n|-|\beta_\star^{(d)}|)=|u^{(d)}|$ and  $\frac{\kappa_j}{\sqrt n}\hat\zeta_j^{(d)}=\frac{\kappa_j}{\sqrt n}(\sqrt n|\hat\beta_j^{(d)}|)^{-\alpha} n^{\alpha/2}$, where $\sqrt n\hat\beta_j^{(d)}=O_p(1)$, then under condition (C4)
		$$\kappa_j\hat\zeta_j^{(d)}\left(\left|\beta_\star^{(d)}+ u^{(d)}/\sqrt n\right|-|\beta_\star^{(d)}|\right)\rightarrow_p |u^{(d)}|\frac{\kappa_j}{\sqrt n} n^{\alpha/2}.$$
	\end{itemize}
	Consequently, it has that $V_n(\bm u)\rightarrow_p V(\bm u)$, where 
	\begin{equation}\label{bound_V}
		V(\bm{u})= \begin{cases}
			\frac{\bm u^{\mathcal A}}{\sqrt n}\sum_{i\in\mathcal I_j}\bm X_i^{\mathcal A}\mathcal L_i^\prime(\bm\beta^{\mathcal A}_\star)+\frac12(\bm u^{\mathcal A})^\top\bm\Phi_j^{\mathcal A}\bm u^{\mathcal A} + O_p\left(1/{\sqrt n}\right) & \text {if~}u^{(d)}=0 \mbox{~if~}\forall d \notin \mathcal{A},  \\
			\infty & \text {otherwise,}
		\end{cases}
	\end{equation}
	with $\bm\Phi_j^{\mathcal A}=\mathbb E \bm X_i^{\mathcal A}(\bm X_i^{\mathcal A})^\top\mathcal L^{\prime\prime}_i(\bm\beta_\star)$ for $i\in\mathcal I_j$. 	
	The unique minimum of $V(\bm u)$ is approximately equal to
	$$\hat{\bm u}=(\hat{\bm u}^{\mathcal A}, \hat{\bm u}^{\mathcal A^c})=\left(\left(\bm\Phi_j^{\mathcal A}\right)^{-1} \frac1{\sqrt n}\sum_{i\in\mathcal I_j}\bm X_i^{\mathcal A}\mathcal L_i^\prime(\bm\beta^{\mathcal A}_\star), \bm 0\right)$$ 
	Following the epi-convergence results of \cite{geyer1994asymptotics} and \cite{knight2000asymptotics}, 
	$$
	\hat{\bm{u}}^{\mathcal{A}}_j \rightarrow_d \left(\mathbf{\Phi}_j^{\mathcal A}\right)^{-1} \mathbf{W}_j^{\mathcal{A}} \quad \text { and } \quad \hat{\bm{u}}_j^{\mathcal{A}^c} \rightarrow_d \mathbf{0} .
	$$
	where $\mathbf{W}_j^{\mathcal{A}}=\frac1{\sqrt n}\sum_{i\in\mathcal I_j}\bm X_i^{\mathcal A}\mathcal L_i^\prime(\bm\beta^{\mathcal A}_\star)\sim N\left(\bm 0, \bm\Psi^\mathcal{A}_j\right)$ with $\bm\Psi_j^{\mathcal{A}}=\mathbb E \bm X_i^\mathcal{A}(\bm X_i^\mathcal{A})^\top\left(\mathcal L_i^\prime(\bm\beta_\star^{\mathcal A})\right)^2$ for $i\in\mathcal I_j$. 
	We can also conclude from (\ref{bound_V}) that 
	$$
	\sqrt n\left(\widehat{\bm\beta}_j^{\mathcal A}-\bm\beta_\star^{\mathcal A}\right) = \hat{\bm u}^{\mathcal A}= \left(\mathbf{\Phi}_j^{\mathcal A}\right)^{-1} \mathbf{W}_j^{\mathcal{A}} +O_p(1/\sqrt n).
	$$
	Note that $\left(\mathbf{\Phi}_j^{\mathcal A}\right)^{-1} \mathbf{W}_j^{\mathcal{A}}=O_p(1)$. 
	
	\hfill $\square$
	
	\noindent\textbf{Proof of Theorem \ref{global_oracle}}.  We first prove result (1) and (3), then (2).
	
	\textit{Proof of result (1).}	
	It is known that $\hat{\sigma}_{j,dd}^2=\sigma_{j,dd}^2+O_p(1/\sqrt{n})$, which results in that 
	$$\frac{\hat w_j^{(d)}}{w_j^{(d)}}=\frac{\hat\sigma_{j,dd}^{-2}}{\sigma_{j,dd}^{-2}}\cdot \frac{\sum_{j=1}^K\sigma_{j,dd}^{-2}}{\sum_{j=1}^K\hat\sigma_{j,dd}^{-2}}=\frac{\sigma_{j,dd}^{-2}+O_p(1/\sqrt n)}{\sigma_{j,dd}^{-2}}\cdot \frac{\sum_{j=1}^K\sigma_{j,dd}^{-2}}{\sum_{j=1}^K\sigma_{j,dd}^{-2}+O_p(K/\sqrt n)}\rightarrow_p 1+O_p(K/\sqrt{n}).$$
	According to the results in Theorem \ref{bound_local_para}, we have that $\mathbb E\left(\hat{\beta}_j^{(d)}-\beta_\star^{(d)}\right)=O_p(1/n)$. 
	As a result, it has that
	\begin{eqnarray}
		&&\mathbb E\|\widehat{\bm\beta}_{\mathrm{WAVE}}-\bm\beta_\star\|_2^2= \mathbb E\sum_{d=1}^p\left(\sum_{j=1}^K  \hat w_j^{(d)}(\widehat{\beta}_{j}^{(d)}-\beta^{(d)}_\star)\right)^2\nonumber \\
		&=&\mathbb E\sum_{d=1}^p\sum_{j=1}^K\sum_{l=1}^K\frac{\hat w_j^{(d)}}{w_j^{(d)}} w_j^{(d)}\frac{\hat w_l^{(d)}}{w_l^{(d)}} w_l^{(d)}(\widehat{\beta}_{j}^{(d)}-\beta^{(d)}_\star)(\widehat{\beta}_{l}^{(d)}-\beta^{(d)}_\star)\nonumber \\
		&=&\left(1+O_p(K/\sqrt{n})\right)\mathbb E\sum_{d=1}^p\sum_{j=1}^K\sum_{l=1}^K w_j^{(d)} w_l^{(d)}(\widehat{\beta}_{j}^{(d)}-\beta^{(d)}_\star)(\widehat{\beta}_{l}^{(d)}-\beta^{(d)}_\star)\nonumber \\
		&=&\left(1+O_p(K/\sqrt{n})\right)\left\{\mathbb E\sum_{d=1}^p\sum_{j=1}^K( w_j^{(d)})^2(\widehat{\beta}_{j}^{(d)}-\beta^{(d)}_\star)^2\right. \nonumber \\
		&&+\left.\mathbb E\sum_{d=1}^p\sum_{j\neq l} w_j^{(d)} w_l^{(d)}(\widehat{\beta}_{j}^{(d)}-\beta^{(d)}_\star)(\widehat{\beta}_{l}^{(d)}-\beta^{(d)}_\star)\right\}.\nonumber 
	\end{eqnarray}
	Since $K=o(N^\xi)$ with $\xi\leq 1/3$, thereby $O_p(K/\sqrt n)=o_p(1)$. Moreover, under the condition $w_j^{(d)}=O_p(1/K)$, it has that 
	$$
	\mathbb E\sum_{d=1}^p\sum_{j=1}^K\sum_{l=1}^K w_j^{(d)}(\widehat{\beta}_{j}^{(d)}-\beta^{(d)}_\star) w_l^{(d)}(\widehat{\beta}_{l}^{(d)}-\beta^{(d)}_\star)=o_p(p/N).
	$$
	Consequently, by proposition \ref{prop1}, we have
	\begin{equation*}
		\begin{aligned}
			&\mathbb E\|\widehat{\bm\beta}_{\mathrm{WAVE}}-\bm\beta_\star\|_2^2=\mathbb E\sum_{d=1}^p\sum_{j=1}^K\left( w_j^{(d)}\right)^2\left(\widehat{\beta}_{j}^{(d)}-\beta^{(d)}_\star\right)^2\leq \frac1K\mathbb E\sum_{d=1}^p\sum_{j=1}^K\left(\widehat{\beta}_{j}^{(d)}-\beta^{(d)}_\star\right)^2\\
			=& \frac1K\sum_{j=1}^K\mathbb E\|\widehat{\bm\beta}_j-\bm\beta_\star\|_2^2
			= \frac{8\kappa_j^2{{\sum_{d=1}^p\left(\zeta_j^{(d)}\right)^2}}K+2{NpB\mu_j}}{N^2c^2}+o_p(p/{N})
		\end{aligned}
	\end{equation*}
	
	\
	
	\textit{Proof of result (3)}. In the following, we prove the asymptotic normality for $\widehat{\bm\beta}_{\mathrm{WAVE}}$. 
	Since it has been proved that $\widehat{\bm\beta}_j^{\mathcal{A}}-{\bm\beta}_\star^{\mathcal{A}}\sim N(\bm 0, n^{-1}\bm\Sigma^{\mathcal A}_j)+O_p(1/n)$ and $\widehat{\bm\beta}_j^{\mathcal{A}^c}\rightarrow_d\bm 0$,  thus
	\begin{eqnarray*}
		\widehat{\bm\beta}_{\mathrm{WAVE}}^\mathcal{A}-\bm\beta^\mathcal{A}_\star&=&\left(\sum_{j=1}^K\left(\widehat{\bm\Lambda}_j^{\mathcal A}\right)^{-1}\right)^{-1}\sum_{j=1}^K\left(\widehat{\bm\Lambda}_j^{\mathcal A}\right)^{-1}\left\{\left(\bm\Psi^{\mathcal A}_j\right)^{-1}\frac1n\sum_{i\in\mathcal I_j}\bm X_i^{\mathcal A}\mathcal L^\prime\left(\bm\beta^{\mathcal A}_\star\right)+O_p(1/n)\right\}\\
		&=&\left(\frac1K\sum_{j=1}^K\left(\bm\Lambda^{\mathcal A}_j\right)^{-1}\right)^{-1}\frac1K\sum_{j=1}^K\left(\bm\Lambda_j^{\mathcal A}\right)^{-1}\left(\bm\Psi^{\mathcal A}_j\right)^{-1}\frac1n\sum_{i\in\mathcal I_j}\bm X_i^{\mathcal A}\mathcal L^\prime(\bm\beta^{\mathcal A}_\star))+O_p(1/n)\\
		&=&\left(\frac1K\sum_{j=1}^K\left(\bm\Lambda^{\mathcal A}_j\right)^{-1}\right)^{-1}\frac1K\sum_{j=1}^K\bm g_j^{\mathcal A} + O_p(1/n),
	\end{eqnarray*}
	where $\bm g_j^{\mathcal A}=\left(\bm\Lambda_j^{\mathcal A}\right)^{-1}\left(\bm\Psi^{\mathcal A}_j\right)^{-1}\frac1n\sum_{i\in\mathcal I_j}\bm X_i^{\mathcal A}\mathcal L^\prime\left(\bm\beta^{\mathcal A}_\star\right)$. The second equality holds because 
	$$\widehat{\bm\Lambda}_j^{\mathcal{A}}=\mbox{diag}\left(\widehat{\bm\Phi}_j^{-1}\widehat{\bm\Psi}_j\widehat{\bm\Phi}_j^{-1}\right)=\bm\Lambda_j^{\mathcal{A}}+O_p(1/\sqrt{n}).$$ 
	According to above results, 
	$$\sqrt n\bm g_j\sim N\left(\bm0,\left(\bm\Lambda_j^{\mathcal A}\right)^{-1}\bm\Sigma_j^{\mathcal A}\left(\bm\Lambda_j^{\mathcal A}\right)^{-1}\right)$$
	Let $\bm e = \bm a\left(\frac1K\sum_{j=1}^K\left(\bm\Lambda^{\mathcal A}_j\right)^{-1}\right)^{-1}$, then under condition (C5) and the Lyapunov central limit theorem, it has that
	\begin{equation*}
		\frac1{\sqrt K}\sum_{j=1}^K\sqrt n\bm e^\top \bm g_j \sim  N\left(\bm 0, \frac1K\bm e^\top \sum_{j=1}^K\left(\bm\Lambda_j^{\mathcal A}\right)^{-1}\bm\Sigma_j^{\mathcal A}\left(\bm\Lambda_j^{\mathcal A}\right)^{-1}\bm e\right)
	\end{equation*}
	Since, 
	\begin{equation*}
		\sqrt N\frac1K\sum_{j=1}^K\bm e^\top \bm g_j=\frac1{\sqrt K}\sum_{j=1}^K\sqrt n\bm e^\top \bm g_j
	\end{equation*}
	We finally get the conclusion that
	\begin{equation*}
		\bm a^\top \left(\widehat{\bm\beta}^{\mathcal A}_{\mathrm{WAVE}}-\bm\beta^{\mathcal A}_\star\right)\sim N\left(\bm 0, \frac1N\bm a^\top \bm V_K^{-1}\bm S_K\bm V_K^{-1} \bm a\right),
	\end{equation*}
	where $ \bm V_K=\frac1K\sum_{j=1}^K\left(\bm\Lambda^{\mathcal A}_j\right)^{-1}$ and $\bm S_K=\frac1K \sum_{j=1}^K\left(\bm\Lambda_j^{\mathcal A}\right)^{-1}\bm\Sigma_j^{\mathcal A}\left(\bm\Lambda_j^{\mathcal A}\right)^{-1}$. 
	
	\
	
	\textit{Proof of result (2).} By taking similar argument like (\ref{feature_consist}), the result can be easily proved.
	
	\hfill $\square$
	
	\textbf{Proof of Corollary \ref{tilde_beta}.} 
	Since 
	\begin{equation*}
		\widetilde{\beta}_{\mathrm{WAVE}}^{(d)}=\mbox{sign}\left(\widehat{\beta}^{(d)}_{\mathrm{WAVE}}\right)\left(\left|\widehat{\beta}^{(d)}_{\mathrm{WAVE}}\right|-\frac{\delta\hat\gamma^{(d)}}{\sum_{j=1}^Kn_j\hat\sigma_{j,dd}^{-2}}\right)_{+},
	\end{equation*}
	Thus, if $d\in\mathcal A$, then $\hat\gamma^{(d)}=|\hat{\beta}_{\mathrm{WAVE}}^{(d)}|^{-\alpha_0}\rightarrow |\beta_\star^{(d)}|^{-\alpha_0}$, so  
	$$\frac{\delta\hat\gamma^{(d)}}{\sum_{j=1}^Kn_j\hat\sigma_{j,dd}^{-2}}=O_p(1/N).$$
	If $d\in\mathcal A^c$, then $\sqrt{N}\hat{\beta}_{\mathrm{WAVE}}^{(d)}=O_p(1)$, so
	$$\frac{\delta\hat\gamma^{(d)}}{\sum_{j=1}^Kn_j\hat\sigma_{j,dd}^{-2}}=\frac{\delta}{\sqrt N}\frac1{(\sqrt N|\hat{\beta}_\mathrm{WAVE}^{d}|)^{\alpha_0}}\frac{N^{(\alpha_0+1)/2}}{\sum_{j=1}^Kn_j\hat\sigma_{j,dd}^{-2}}=O(1)\frac{\delta}{\sqrt N}N^{(\alpha_0-1)/2}\rightarrow \infty,$$
	where the above conclusion holds because the fact  $\sum_{j=1}^Kn_j\hat\sigma_{j,dd}^{-2}=O(N)$ and condition (C6).
	
	\hfill $\square$
	
	
	\bibliographystyle{apalike}
	\bibliography{bib_junlu_abbrev}

\end{document}